\documentclass[useAMS,usenatbib,usegraphicx]{mn2e}

\title[A search for optical bursts from the RRAT J1819--1458]{A search for optical bursts from the rotating radio transient J1819--1458 with ULTRACAM}
\author[V. S. Dhillon et al.]{V. S. Dhillon$^{1}$\thanks{E-mail:
vik.dhillon@sheffield.ac.uk}, T. R. Marsh$^{2}$ and S. P. Littlefair$^{1}$\\
$^{1}$Department of Physics and Astronomy, University of Sheffield, 
Sheffield S3 7RH, UK \\
$^{2}$Department of Physics, University of Warwick, Coventry CV4 7AL, UK}
\begin{document}

\date{Submitted for publication in the Monthly Notices of the Royal
  Astronomical Society on 2006 May 9.}

\pagerange{\pageref{firstpage}--\pageref{lastpage}} \pubyear{2006}

\maketitle

\label{firstpage}

\begin{abstract}

  We report on the search for optical bursts from J1819--1458, a
  member of the recently discovered Rotating Radio Transients (RRATs).
  J1819--1458 exhibits 3 millisecond bursts with a peak flux of
  $f_{\nu}^{1.4 GHz}$ = 3.6 Jy every $\sim 3.4$ minutes, implying that
  it is visible for only $\sim 1$ second per day at radio
  wavelengths. Assuming that the optical light behaves in a similar
  manner, the most sensitive way of detecting RRATs is hence not to
  take long exposures of the field, but instead to capture individual
  bursts using a high-speed camera mounted on a large aperture
  telescope. Using ULTRACAM on the 4.2-m William Herschel Telescope
  (WHT) we obtained 97\,100 images of the field of J1819--1458, each
  of 18.1 milliseconds exposure time and with essentially no dead-time
  between the frames. We find no evidence for bursts in $u'$, $g'$ and
  $i'$ at magnitudes brighter than 15.1, 17.4 and 16.6 (5$\sigma$),
  corresponding to fluxes of less than 3.3, 0.4 and 0.8 mJy at
  3560\AA, 4820\AA\ and 7610\AA, respectively.

\end{abstract}


\section{Introduction}

The RRATs are a remarkable new class of variable star characterized by
their radio bursts of duration 2--30 milliseconds which recur every
4--180 minutes \citep{mclaughlin06}. The RRATs, of which 11 are
currently known, exhibit periodicities of 0.4--7 seconds, inferred by
dividing the intervals between bursts by the largest common
denominator. Such periods are long in comparison with most radio
pulsars and are instead reminiscent of the periods found in the
radio-quiet Anomalous X-ray Pulsars (AXPs), Soft Gamma Repeaters
(SGRs) and X-ray Dim Isolated Neutron Stars (see \citealt{woods06} and
\citealt{haberl04}). The distances to the RRATs can be estimated from
their dispersion measures and it is found that they lie 2--7 kpc away
in the Galactic plane. In the three RRATs with the most frequent
bursts it has also been possible to measure a period derivative,
showing no evidence for binarity but instead that these objects spin
down like other pulsars. In the case of J1819--1458, the inferred
magnetic field strength is high ($B\sim 5 \times 10^{13}$ G),
providing another link between the RRATs and the magnetars (i.e. the
AXPs and SGRs).

The sporadic nature of the bursts in RRATs makes localization to
better than the 14 arcminute beam of the Parkes Telescope
difficult. Fortunately, the positions of the three RRATs with period
derivatives can be refined through radio timing and were quoted to an
accuracy of arcseconds by \citet{mclaughlin06}. This enabled
\cite{reynolds06} to identify the {\em Chandra} source CXOU
J181934.1--145804, lying within 2 arcseconds\footnote{It should be
  noted that this assumes updated radio coordinates which differ from
  the radio position listed in \citealt{mclaughlin06} (Steve Reynolds,
  private communication).} of the radio source, as the X-ray
counterpart to J1819--1458. The X-ray properties of the source, which
is point-like and shows no variability, are consistent with thermal
emission from a cooling neutron star \citep{reynolds06}, lending
further weight to the hypothesis that RRATs are rotating neutron
stars.

To further constrain the nature of RRATs, it is desirable to observe
them at different wavelengths. \cite{reynolds06} made a first attempt
at this by searching optical and infrared archives for counterparts to
CXOU J181934.1--145804. They found none, but their limits are not
particularly deep ($I=17.5, J=15.6, H=15.0, K=14.0$). Taking longer
exposures to go deeper is not necessarily the best solution, however,
as the RRATs may have very faint persistent optical/IR emission and
only emit strongly at these wavelengths during
bursts\footnote{\cite{reynolds06} found no evidence for X-ray bursts
  in CXOU J181934.1--145804, but this does not mean optical bursts will
  be undetectable -- the AXP 4U\,0142+61, for example, exhibits a
  pulsed fraction 5--7 times greater in the optical than the X-ray
  \citep{dhillon05}.  In addition, by analogy with the Crab pulsar,
  searching for optical bursts (as opposed to any persistent emission)
  may also be the most sensitive method of detecting RRATs -- the main
  optical pulse of the Crab is 5 magnitudes brighter in the B and
  V-bands compared to its persistent light level and is coincident (to
  within 100 $\mu$s) with the radio peak \citep{golden00}.}. Given that
the duration of the bursts in J1819--1458, for example, only total
$\sim 1$ second a day, the best strategy would then be to reduce the
contribution of the sky and take a continuous sequence of extremely
short exposures on a large-aperture telescope covering a number of
burst cycles in order to catch a burst in one or two of the frames.

With the high-speed, triple-beam CCD camera ULTRACAM at our disposal
\citep{dhillon01}, we have the ideal tool with which to search for
optical bursts from RRATs. In this paper, we report on an attempt to
detect such bursts from the RRAT J1819--1458. We selected this RRAT as
it exhibits both the most frequent and most powerful radio bursts -- a
3 millisecond burst of 3.6 Jy (at 1400 MHz) occurring every $\sim 3.4$
minutes, with the burst intervals showing a periodicity of 4.26
seconds. Thanks to the X-ray identification by \cite{reynolds06},
J1819--1458 is also the RRAT with the best-determined position,
accurate to 0.5 arcseconds, and is one of the closer RRATs to the
Earth (at a distance of 3.6 kpc).

\section{Observations and data reduction}
\label{obsred}

The observations of J1819--1458 presented in this paper were obtained
with ULTRACAM (\citealt{dhillon01}, \citealt{beard02}) at the
Cassegrain focus of the 4.2-m William Herschel Telescope (WHT) on La
Palma.  ULTRACAM is a CCD camera designed to provide imaging
photometry at high temporal resolution in three different colours
simultaneously. The instrument provides a 5 arcminute field on its
three $1024\times1024$ E2V 47-20 CCDs (i.e. 0.3
arcseconds/pixel). Incident light is first collimated and then split
into three different beams using a pair of dichroic beamsplitters. For
the observations presented here, one beam was dedicated to the SDSS
$u'$ filter, another to the SDSS $g'$ filter and the third to the SDSS
$i'$ filter. Because ULTRACAM employs frame-transfer chips, the
dead-time between exposures is negligible: we used ULTRACAM in its
highest-speed {\em drift mode}, with 2 windows each of $60\times60$
pixels and $2\times2$ binning, resulting in an exposure time of 18.1
milliseconds and a dead-time of 1.4 milliseconds. A total of 97\,100
consecutive frames of J1819--1458 were obtained from 06:07--06:38 UTC
on 2006 March 11, with each frame time-stamped to a relative accuracy
of better than 50~$\mu$s using a dedicated GPS system.  The data were
obtained in photometric conditions, with no moon and $i'$-band seeing
of $\sim 1.5$ arcseconds.

\begin{figure*}
\centering
\includegraphics[width=3.5cm,angle=270]{finding_chart.ps}~\includegraphics[width=2.0cm,angle=270]{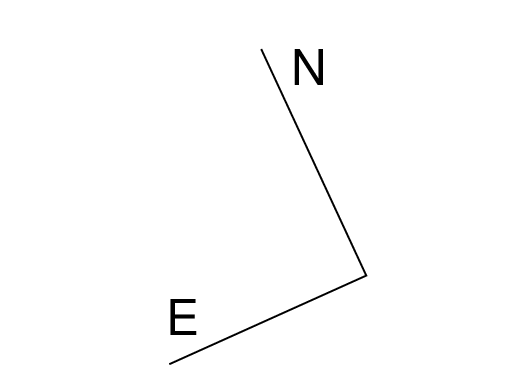}
\caption{Image of the field of J1819--1458 in the $i$-band, taken by
  summing 9 acquisition frames with a total exposure time of 30
  seconds.  The two ULTRACAM windows used to acquire high-speed data
  on J1819--1458 are shown by the dashed boxes. The X-ray position of
  J1819--1458 derived by \protect\cite{reynolds06} is marked in the
  left-hand window. The plate scale is 0.3 arcseconds/pixel and the
  orientation of the field is shown on the right-hand side.}
\label{finding_chart}
\end{figure*}

Figure~\ref{finding_chart} shows part of an ULTRACAM $i'$-band
acquisition image of the field of J1819--1458 with the 2 drift-mode
windows and X-ray position marked. The data were reduced using the
ULTRACAM pipeline data reduction system. In order to perform aperture
photometry, we had to determine the pixel position of J1819--1458 on
the ULTRACAM detectors. This was achieved by offsetting from the star
at pixel position (282, 44) in figure~\ref{finding_chart}, which we
identified as 2MASS\,18193423--1457589 ($\alpha_{2000}$: 18 19 34.23,
$\delta_{2000}$: --14 57 59.0), to the X-ray position of J1819--1458
given by \cite{reynolds06}. Given the uncertainty in the 2MASS
position (0.1 arcseconds), the {\em Chandra} position (0.5 arcseconds)
and the plate scale (0.01 arcseconds/pixel), we estimate that the
resulting error in the location of J1819--1458 on the ULTRACAM
detectors is $\sim 0.6$ arcseconds. Due to this uncertainty, we
extracted light curves using a series of software apertures centred on
the X-ray position of J1819--1458, with radii ranging from 0.9 to 5.4
arcseconds and with the sky level determined from an annulus
surrounding the largest aperture.

\section{Results}

\begin{figure*}
\centering
\includegraphics[width=9cm,angle=270]{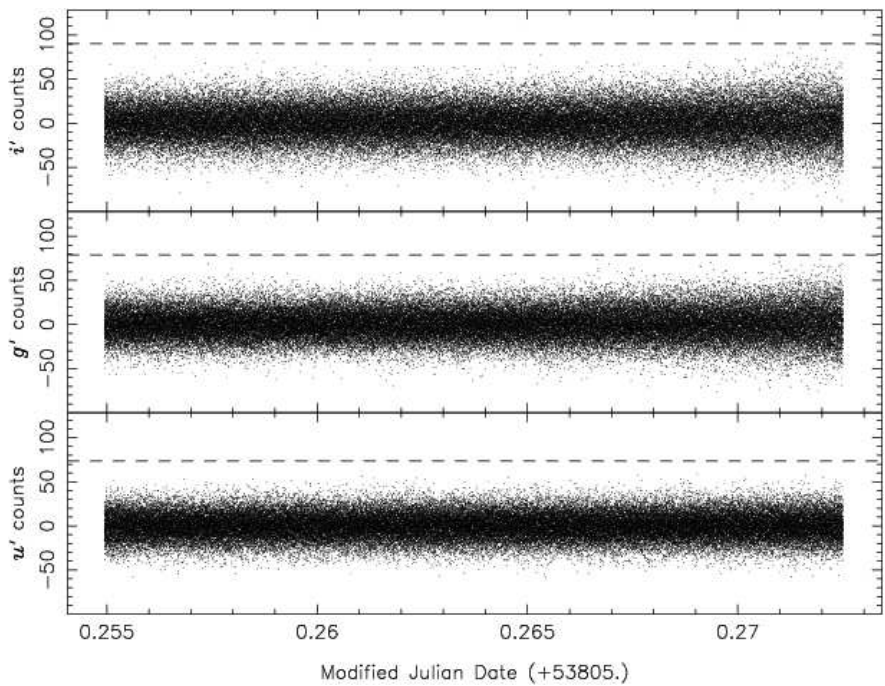}
\caption{From top to bottom, $i'$, $g'$ and $u'$ light curves of
  J1819--1458, extracted using an aperture radius of 1.8 arcseconds.
  The dashed lines show the level above which points would deviate by
  more than +5$\sigma$ from the mean. Note the increased scatter
  towards the end of the run, particularly evident in $i'$, due to the
  approach of morning twilight. As a result, this part of the light
  curve was omitted during the standard deviation calculation.}
\label{light_curve}
\end{figure*}

Figure~\ref{light_curve} shows the $u'$, $g'$ and $i'$ light curves of the
field around J1819--1458 for an aperture radius of 1.8 arcseconds. The
signature of an optical counterpart to the radio bursts detected by
\cite{mclaughlin06} would be a series of deviant points lying greater
than, say\footnote{We have chosen 5$\sigma$ as we have $\sim10^5$
data points and only one point in $\sim10^6$ would be expected to be
greater than 5$\sigma$ from the mean in a Gaussian distribution.},
5$\sigma$ from the mean. Given that the exposure time is significantly
longer than the radio burst duration, one would expect only one or two
points per burst. Furthermore, given the mean interval between radio
bursts and the length of the ULTRACAM dataset, one would expect there
to be approximately 9 such deviant points in the light curve, each
separated by $\sim 3.4$ minutes.

The dashed line in figure~\ref{light_curve} shows the +5$\sigma$
deviation level. It can be seen that there are no points lying above
this line. The light curves derived from larger aperture radii do show
1--2 points lying well above the +5$\sigma$ line, but a visual
inspection of the corresponding images revealed that the events were
all due to cosmic rays (as shown by their non-stellar profiles and
the fact that they were only visible in one of the three CCD
chips). The implication is that we have not detected any evidence for
optical counterparts to the radio bursts from J1819--1458.

\begin{table}
\centering
\caption{Limiting magnitudes (5$\sigma$) for the optical bursts from J1819--1458
  as a function of the extraction aperture radius.}
\begin{tabular}{cccc}
 & & & \\
\hline
Aperture radius & \multicolumn{3}{c}{Limiting magnitude (5$\sigma$)} \\
(arcseconds) & $u'$ & $g'$ & $i'$ \\
0.9 & 16.0 & 18.3 & 17.5 \\
1.8 & 15.1 & 17.4 & 16.6 \\
2.7 & 14.6 & 16.8 & 16.0 \\
3.6 & 14.1 & 16.4 & 15.6 \\
4.5 & 13.7 & 16.0 & 15.2 \\
5.4 & 13.4 & 15.6 & 14.8 \\
\hline
\end{tabular}
\label{limiting_magnitudes}
\end{table}

Although a negative result, it is useful to place a magnitude limit on
the optical bursts from J1819--1458 in order to inform future optical
studies of RRATs. The magnitude limits are shown in
table~\ref{limiting_magnitudes} and can be seen to depend on the
aperture radius used for the extraction, as a burst has to be brighter
to be visible at the 5$\sigma$ level in a larger aperture due to the
larger number of pixels in the aperture and hence the increased noise
from the sky and detector. A reasonable choice of aperture would be
1.8 arcseconds, as this corresponds to 3$\sigma$ from the expected
position of J1819--1458 (see section~\ref{obsred}). In this case, we
are able to say that J1819--1458 shows no evidence for optical bursts
brighter than 15.1, 17.4 and 16.6 in $u'$, $g'$ and $i'$,
respectively. This corresponds to $U$, $B$ and $I$ magnitudes of 15.4,
17.9 and 16.1, respectively, where we have used the conversion
equations of \cite{smith02} and assumed $r' = i'$. 

The corresponding flux limits are $<3.3$, 0.4 and 0.8 mJy in $u'$,
$g'$ and $i'$, respectively, calculated using equation 2 of
\cite{fukugita96}. It is not straightforward to assign a wavelength to
each of these fluxes, as the effective wavelengths of the SDSS filters
can be defined in a number of ways (see \citealt{fukugita96}) and depend
on a number of parameters, including the throughput of the optics, the
detector quantum efficiencies, the atmospheric extinction and the
spectrum of the star. Moreover, there is a systematic uncertainty in
the conversion of SDSS magnitude to flux of 5--10\% \citep{smith02}.
We calculate effective wavelengths for the $u'$, $g'$ and $i'$ fluxes
quoted above of 3560\AA, 4820\AA\ and 7610\AA, respectively, assuming
a flat spectral source.

Taking our most sensitive flux limit of $<0.4$ mJy at $g'$
($6.2\times10^{14}$ Hz), we can compare this to the radio flux of 3600
mJy at $1.4\times10^9$ Hz measured by \cite{mclaughlin06} to deduce
that the spectral slope must be steeper than $f_{\nu}\propto
\nu^{-0.7}$. For comparison, the radio-to-optical slope of the Crab
pulsar is $\sim -0.2$ \citep{lyne05b}, but it is not straightforward to
interpret such coarse spectral energy distributions because the
optical and radio photons have very different origins (the optical is
incoherent emission from the outer gap, the radio is coherent emission
from both the polar cap and the outer gap -- see \cite{lyne05b} and
references therein).

It should be noted that the dead-time of ULTRACAM during our
J1819--1458 observations was less than half the duration of the radio
bursts observed by \cite{mclaughlin06}. This makes it unlikely we
missed a single optical burst whilst ULTRACAM was reading out, let
alone the expected 9 bursts. Moreover, if the emission mechanisms in
J1819--1458 are similar to those of the Crab pulsar, in which the
optical pulse is known to be approximately 5 times wider than the
radio pulse \citep{shearer03}, it makes it even more unlikely that our
non-detection of optical pulses is due to the dead-time of ULTRACAM.

We searched for periodicities in the light curve shown in
figure~\ref{light_curve} using a Lomb-Scargle periodogram
\citep{press89}.  No evidence for a significant peak around the
proposed 4.26 second spin period was found.

\section{Discussion}

We find no evidence for optical analogues to the radio bursts seen in
the Rotating Radio Transient J1819--1458. Using a frame rate of
$\sim50$~Hz, ULTRACAM has enabled us to place 5$\sigma$ limits on the
burst magnitudes of 15.1, 17.4 and 16.6 in $u'$, $g'$ and $i'$,
respectively. In comparison with the AXP 4U\,0142+61, for example,
which has magnitudes of $i'=23.7$, $g'=27.2$ and $u'>25.8$
\citep{dhillon05}, our limits on J1819--1458 do not appear to be
particularly deep. To place our limits in some context, therefore, it
should be noted that if we had taken a single 1 hour exposure of the
field with the WHT under identical conditions, and assuming the object
emitted 18 bursts, each of $i'=16.6$ and 18.1 milliseconds duration,
we would have detected the object at only $\sim 0.7\sigma$.  Using the
high-speed photometry technique described in this paper, on the other
hand, we would have detected the source at 5$\sigma$. The difference
in sensitivity between the two techniques is due to the fact that the
long exposures would be sky limited, whereas the data presented in
this paper are readout-noise limited.

Due to the nature of the bursts in RRATs, therefore, the only way we
can significantly improve on the magnitude limits is to observe at
higher frame rates (in order to reduce the small contribution of sky
noise still further) and/or use a larger aperture telescope (in order
to increase the number of counts detected from each burst). The
discussion above assumes, of course, that the optical and radio light
behave in a similar manner. If, however, the optical light has only a
low (or no) pulsed fraction, then deep, long-exposure imaging might
prove fruitful, as might searches for pulsed light on the proposed
spin period of the neutron star (e.g. \citealt{dhillon05}).

\section*{Acknowledgments}

TRM acknowledges the support of a PPARC Senior Research Fellowship.
SPL is supported by PPARC grant PPA/G/S/2003/00058. ULTRACAM is
supported by PPARC grant PP/D002370/1.  The William Herschel Telescope
is operated on the island of La Palma by the Isaac Newton Group in the
Spanish Observatorio del Roque de los Muchachos of the Instituto de
Astrof\'{i}sica de Canarias. We thank the anonymous referee for
comments which significantly improved the presentation of the results.

\bibliographystyle{mn2e}
\bibliography{abbrev,refs}

\begin{thebibliography}{}

\bibitem[\protect\citeauthoryear{Beard, Vick, Atkinson, Dhillon, Marsh, McLay,
  Stevenson \& Tierney}{Beard et~al.}{2002}]{beard02}
Beard S.~M.,  Vick A.~J.~A.,  Atkinson D.,  Dhillon V.~S.,  Marsh T.~R.,  McLay
  S.,  Stevenson M.~J.,    Tierney C.,  2002, in Lewis H.,  ed., Advanced
  Telescope and Instrumentation Control Software II. SPIE, 4848, p.~218

\bibitem[\protect\citeauthoryear{{Dhillon} \& {Marsh}}{{Dhillon} \&
  {Marsh}}{2001}]{dhillon01}
{Dhillon} V.~S.,  {Marsh} T.~R.,  2001, New Astronomy Review, 45, 91

\bibitem[\protect\citeauthoryear{{Dhillon}, {Marsh}, {Hulleman}, {van
  Kerkwijk}, {Shearer}, {Littlefair}, {Gavriil} \& {Kaspi}}{{Dhillon}
  et~al.}{2005}]{dhillon05}
{Dhillon} V.~S.,  {Marsh} T.~R.,  {Hulleman} F.,  {van Kerkwijk} M.~H.,
  {Shearer} A.,  {Littlefair} S.~P.,  {Gavriil} F.~P.,    {Kaspi} V.~M.,  2005,
  MNRAS, 363, 609

\bibitem[\protect\citeauthoryear{Fukugita, Ichikawa, Gunn, Doi, Shimasaku \&
  Schneider}{Fukugita et~al.}{1996}]{fukugita96}
Fukugita M.,  Ichikawa T.,  Gunn J.~E.,  Doi M.,  Shimasaku K.,    Schneider
  D.~P.,  1996, AJ, 111, 1748

\bibitem[\protect\citeauthoryear{{Golden}, {Shearer}, {Redfern}, {Beskin},
  {Neizvestny}, {Neustroev}, {Plokhotnichenko} \& {Cullum}}{{Golden}
  et~al.}{2000}]{golden00}
{Golden} A.,  {Shearer} A.,  {Redfern} R.~M.,  {Beskin} G.~M.,  {Neizvestny}
  S.~I.,  {Neustroev} V.~V.,  {Plokhotnichenko} V.~L.,    {Cullum} M.,  2000,
  A\&A, 363, 617

\bibitem[\protect\citeauthoryear{Haberl}{Haberl}{2004}]{haberl04}
Haberl F.,  2004, Mem.~Soc.~Astron.~Ital., 75, 454

\bibitem[\protect\citeauthoryear{Lyne \& Graham-Smith}{Lyne \&
  Graham-Smith}{2005}]{lyne05b}
Lyne A.~G.,  Graham-Smith F.,  2005, Pulsar Astronomy.
Cambridge University Press, Cambridge

\bibitem[\protect\citeauthoryear{McLaughlin, {Lyne}, {Lorimer}, {Kramer},
  {Faulkner}, {Manchester}, {Cordes}, {Camilo}, {Possenti}, {Stairs}, {Hobbs},
  {D'Amico}, {Burgay} \& {O'Brien}}{McLaughlin et~al.}{2006}]{mclaughlin06}
McLaughlin M.~A.,  {Lyne} A.~G.,  {Lorimer} D.~R.,  {Kramer} M.,  {Faulkner}
  A.~J.,  {Manchester} R.~N.,  {Cordes} J.~M.,  {Camilo} F.,  {Possenti} A.,
  {Stairs} I.~H.,  {Hobbs} G.,  {D'Amico} N.,  {Burgay} M.,    {O'Brien} J.~T.,
   2006, Nat, 439, 817

\bibitem[\protect\citeauthoryear{Press \& Rybicki}{Press \&
  Rybicki}{1989}]{press89}
Press W.~H.,  Rybicki G.~B.,  1989, ApJ, 338, 277

\bibitem[\protect\citeauthoryear{{Reynolds}, {Borkowski}, {Gaensler}, {Rea},
  {McLaughlin}, {Possenti}, {Israel}, {Burgay}, {Camilo}, {Chatterjee},
  {Kramer}, {Lyne} \& {Stairs}}{{Reynolds} et~al.}{2006}]{reynolds06}
{Reynolds} S.~P.,  {Borkowski} K.~J.,  {Gaensler} B.~M.,  {Rea} N.,
  {McLaughlin} M.,  {Possenti} A.,  {Israel} G.,  {Burgay} M.,  {Camilo} F.,
  {Chatterjee} S.,  {Kramer} M.,  {Lyne} A.,    {Stairs} I.,  2006, ApJ, 639,
  L71

\bibitem[\protect\citeauthoryear{{Shearer}, {Stappers}, {O'Connor}, {Golden},
  {Strom}, {Redfern} \& {Ryan}}{{Shearer} et~al.}{2003}]{shearer03}
{Shearer} A.,  {Stappers} B.,  {O'Connor} P.,  {Golden} A.,  {Strom} R.,
  {Redfern} M.,    {Ryan} O.,  2003, Sci, 301, 493

\bibitem[\protect\citeauthoryear{Smith, Tucker, Kent, Richmond, Fukugita,
  Ichikawa, Ichikawa, Jorgensen, Uomoto, Gunn, Hamabe, Watanabe, Tolea, Henden,
  Annis, Pier, McKay, Brinkmann, Chen, Holtzman, Shimasaku \& York}{Smith
  et~al.}{2002}]{smith02}
Smith J.~A.,  Tucker D.~L.,  Kent S.,  Richmond M.~W.,  Fukugita M.,  Ichikawa
  T.,  Ichikawa S.,  Jorgensen A.~M.,  Uomoto A.,  Gunn J.~E.,  Hamabe M.,
  Watanabe M.,  Tolea A.,  Henden A.,  Annis J.,  Pier J.~R.,  McKay T.~A.,
  Brinkmann J.,  Chen B.,  Holtzman J.,  Shimasaku K.,    York D.~G.,  2002,
  AJ, 123, 2121

\bibitem[\protect\citeauthoryear{Woods \& Thompson}{Woods \&
  Thompson}{2006}]{woods06}
Woods P.~M.,  Thompson C.,  2006, in Lewin W. H.~G.,  van~der Klis M.,  eds,
  Compact Stellar X-ray Sources. CUP, Cambridge, in press (astro-ph/0406133)

\end{thebibliography}

\label{lastpage}

\end{document}